\documentclass[aps,twocolumn,superscriptaddress]{revtex4}

\usepackage{graphicx}
\usepackage{amsmath}
\usepackage{amsfonts}
\usepackage{amssymb}

\begin{document}

\def\be{\begin{equation}}
\def\ee{\end{equation}}
\def\bea{\begin{eqnarray}}
\def\eea{\end{eqnarray}}
\def\bma{\begin{mathletters}}
\def\ema{\end{mathletters}}
\newcommand{\one}{\mbox{$1 \hspace{-1.0mm}  {\bf l}$}}
\newcommand{\eins}{\mbox{$1 \hspace{-1.0mm}  {\bf l}$}}
\def\C{\hbox{$\mit I$\kern-.7em$\mit C$}}
\newcommand{\tr}{{\rm tr}}
\newcommand{\half}{\mbox{$\textstyle \frac{1}{2}$}}
\newcommand{\shalf}{\mbox{$\textstyle \frac{1}{\sqrt{2}}$}}
\newcommand{\ket}[1]{ | \, #1  \rangle}
\newcommand{\bra}[1]{ \langle #1 \,  |}
\newcommand{\proj}[1]{\ket{#1}\bra{#1}}
\newcommand{\kb}[2]{\ket{#1}\bra{#2}}
\newcommand{\bk}[2]{\langle \, #1 | \, #2 \rangle}
\def\II{I(\{p_k\},\{\rho_k\})}
\def\ss{{\cal K}}
\tolerance = 10000

\bibliographystyle{apsrev}

\title{Fermionizing a small gas of ultracold bosons}

\author{B. Paredes}
\affiliation{Max--Planck Institute for Quantum Optics,
Garching, Germany}
\author{P. Zoller}
\affiliation{Institute for Theoretical Physics, University of
Innsbruck, Austria}
\author{J. I. Cirac}
\affiliation{Max--Planck Institute for Quantum Optics,
Garching, Germany}

\begin{abstract}
We study the physics of a rapidly--rotating gas of
ultracold atomic bosons, with an internal degree of freedom. 
We show that in the  limit of rapid rotation of the trap the  problem 
 exactly maps onto that of non--interacting fermions with spin 
in the lowest Landau level. The spectrum of
the real bosonic system is identical to the one of the
effective fermions, with the same eigenvalues and the
same density of states. 
When the ratio of the number of atoms to the spin degeneracy
is an integer number, the ground state for the effective fermions
is an integer quantum Hall state. The corresponding bosonic
state is a  fractional quantum Hall liquid whose filling factor ranges
in the sequence $\nu=1/2, 2/3, 3/4, \ldots$, as the spin degeneracy
increases. Anyons with
$1/2,1/3,1/4, \ldots$ statistics can be created by inserting lasers
with the appropriate polarizations. 
A special situation appears when  the spin
degeneracy equals the number of atoms in the gas. The ground state is 
then the  product of a completely antisymmetric
spin state and a $\nu=1$ Laughlin state. In this case
the system exhibits fermionic
excitations with fermionic statistics though the real components are
bosonic atoms.

\end{abstract}

\date{\today}
\pacs{PACS} \maketitle

\section{Introduction}

Ultra cold atomic gases have been shown to be
ideal candidates to observe several
intrinsically  quantum phenomena,
since they can be easily controlled and manipulated
by electromagnetic fields.
What type of quantum phenomena  we can observe depends
dramatically on the strength of the atomic interaction,
going from one particle phenomena to highly entangled quantum states.

The first experimental highlight with these systems 
was the achievement of
Bose-Einstein condensation of weakly interacting
atomic gases \cite{BEC}. This success motivated
a lot of experimental and theoretical work with cold
bosons in the limit of weak interactions. In this regime 
atoms  essentially occupy the same single
particle state, and the physics
of the system is perfectly understood via a
Gross-Pitaievski equation for the macroscopic 
wave function. Many interesting phenomena have been
studied in this regime and there still exist certain
open questions  \cite{Stringari}.

On the other hand, it seems clear that
 going in the direction of strongly correlated
atoms, and exploring what happens beyond the validity
of mean field theory, promises to open a new path to 
the observation of novel physical phenomena.
Strong correlations between particles have been
shown to lead to unusual fascinating effects in condensed
matter systems.
In the same way, if an atomic system
is  driven   into the
 strongly interacting regime,
similar interesting phenomena  may be observed.
Moreover, since atomic gases can be well controlled and
manipulated, they  may  
provide us  with new scenarios for the observation
of these phenomena and, perhaps, new ones that have not
been accessible  so far
in solid state environments.
The difficulty here is, however, how to drive an atomic
system into a strongly interacting regime.
In order to enter this regime one should make the typical
interaction energy larger than any other (single particle)
energy scale in the problem. 
Naively speaking, it may seem that the easiest way to do this would be
just to increase the interaction strength, either by
increasing the scattering length or the atomic density.
However, though the scattering length can be effectively enlarged
by using Feshbach resonances \cite{Feshbach}, and the atoms squeezed to
increase the density, both methods may lead to  undesirable
instabilities in the system.

Alternatively, in order to make interactions to  dominate, one can decrease
the single particle energies by creating degeneracies.
Here, one possibility is to use internal levels,
as  in the recent proposal to entangle atomic beams
by generation of spin-squeezed states \cite{Sorensen}.
Degeneracies can
 also be induced by using a periodic
external potential, since  single particle states corresponding to
different
wells will have the same energy.
In the limit in which the  amplitude of tunneling  between wells is small
compared to  the typical interaction energy, the system
has been predicted to undergo a transition from a superfluid phase
to a Mott insulator phase \cite{Dieter}, which is highly correlated.
This quantum  phase transition has been recently achieved 
experimentally
in an optical lattice \cite{Bloch}.
Another way of achieving a quasi-degeneracy in the atomic motional
states is to rotate  the trap that confines the atoms. 
When the frequency of rotation is small
compared to the frequency of the trap, the system
forms a vortex condensate, with a number of vortices 
that increases as the rotation becomes faster \cite{vortices}.
This vortex condensate is a non--correlated state,
in which all the atoms occupy the same single particle
quantum state.
 However, when the rotation 
frequency is close to the harmonic trap frequency 
 the system has been predicted to
enter a {\em fractional quantum Hall regime}, in which the
atoms are strongly correlated \cite{otros, Belen}. 
The ground state is a $\frac{1}{2}$--Laughlin state,
a highly entangled liquid with nearly uniform density.
This atomic liquid offers the fascinating
possibility of creating and manipulating anyons with
$\frac{1}{2}$--fractional--statistics \cite{Belen}.
By piercing the system with  off resonant localized lasers,
 $\frac{1}{2}$--Laughlin quasiholes may be
created and manipulated to probe directly their
fractional statistics.
For the  fractional quantum Hall regime to be  
experimentally achieved in current rotating traps
one would need bosonic gases 
consisting of a small number of atoms,
a limit that has  remained practically
unexplored. A particularly interesting exception is
a  recent experiment with an optical
lattice \cite{Bloch}, that has   achieved a regime in which a few number of
atoms are confined 
in each well of the lattice.
Since the two-dimensional rotating realization of this lattice 
may be accessible in near future experiments,  further theoretical
investigation in the physics of small rotating bosonic gases is
highly motivated.

In this paper we extend the results of our previous work \cite{Belen} on
a rapidly rotating gas of bosonic atoms, to the case in which the
atoms have an internal degree of freedom. We will show
that the addition of internal levels
introduces new aspects of the problem,
and novel exotic correlated  phases come out.

The problem of strongly interacting bosons with spin
in a rapidly rotating trap 
is, in principle, very difficult
to tackle; in particular it is not easy to guess 
what type of ground states and excitations  appear.
In this work we present a formalism,
what we call {\em fermionization} of the bosonic
system, within
which the problem takes a very simple form.
We will show that in the limit of rapid rotation and
small temperatures,
the problem of interacting bosons maps exactly 
onto a problem of non-interacting fermions.
The ground state  corresponds to a
Fermi sea in which the fermions fill up the single
particle states in order of increasing energy.
Furthermore, the lowest energy excitations  consist
of fermion-hole excitations in the vicinity of
the Fermi level.
The spectrum of the real bosonic system is identical
to the one of the effective non-interacting fermions,
with the same eigenvalues and the same density of states.

The fermionization scheme is valid for any number 
of particles $N$ and any number of internal states $n$,
provided that the atomic interaction does not depend
on the internal states of the atoms.
Perhaps the most interesting states appear whenever the
ratio $N/n$ is an integer number. The effective fermions then
form an integer quantum Hall state with filling factor
$\nu=n$ \cite{QHE}. The corresponding bosonic state is a  fractional
quantum Hall liquid  with filling factor $\nu=n/(1+n)$.
This fractional liquid is a multicomponent state 
made out of $n$ copies of a $\frac{1}{1+n}$--Laughlin--liquid \cite{Laughlin},
one for each internal state.
Thus, by increasing the number of internal states we can go 
beyond $\nu=1/2$, and lead the system into new
fractions in the sequence $\nu=1/2,2/3,3/4, \ldots$.  All type of anyons
 \cite{anyons}
with statistics $p/q$, $ p<q$ can now be created by choosing the
adequate spin degeneracy and inserting lasers 
with a polarization and frequencies 
which couple to a given atomic internal state (spin component). 

An exotic situation appears when the number of
internal states equals the number of atoms in the
trap. The effective fermions form in this case
a pure spinor state with no spatial dependence.
The corresponding bosonic state is then
the product of  a completely antisymmetric spin state
and a $\nu=1$ Laughlin state, which is fully antisymmetric too.
This means  that as long as the spin degrees of freedom
are not excited the behaviour of the system
will be fermionic. We will show that the elementary
excitations of the systems are true fermions
 with fermionic statistics,
though 
the real components of the system are bosons.

The formalism of fermionization that we present in this
work can be also applied to explore several
 interesting possibilities.
A very attractive one consists of inducing
pairing in the effective fermions, so that they form
a superconducting state.
We will show that this is in  fact possible when the
atomic interaction  depends on the internal state.
We will consider the case of atoms with two internal 
levels when the interaction between atoms in the same spin state 
is  much stronger than the one between
atoms in a different spin state.
We will see that the effective fermions configurate
in pairs, forming a paired state identical to the BCS
state of electrons with spin triplet p--wave pairing.
The corresponding bosonic state, a Pfaffian state,
is very well known in the
context of the
quantum Hall effect \cite{Read}.

This paper is organized as follows.
In section II we develop the fermionization scheme that maps the
problem of interacting bosons in a rapid rotating trap onto one
 of non-interacting fermions. Using  this formalism we show in section
III that, when the ratio $N/n$ is an integer number, the ground
state of the system is a $\nu=n/(1+n)$ fractional quantum Hall
liquid.  Starting with these correlated liquids, we show how to
create anyons with $p/(1+n)$ fractional
statistics, with $1 \leq p \leq n$. We also discuss a possible experiment
to reveal directly the statistics of anyons  in a Ramsey-type 
interferometer. In section IV we study the  special situation in which $N=n$.
In section V we illustrate the results of the preceeding sections
with some numerical exact diagonalizations for $N=4,6,8$ particles.
The experimental conditions required to observe the fractional
quantum Hall states and the anyons described in this work are
analyzed in section VI. As an example of the interesting 
possibilities of the fermionization scheme we show in section VII
how pairing in the effective fermions can be induced in a system
of atoms with two internal levels.
Finally, in section VIII we summarize  the results of this work,
and discuss a possible scenario to observe the fractional quantum
Hall states and the anyons.

\section{The fermionization} 

We consider a set of $N$ bosonic atoms confined in a potential
which rotates in the $x-y$ plane at a frequency $\Omega$. 
The atoms
have an internal degree of freedom, that we will
call spin, which can take $n=2s+1$
values.
We will 
assume that the confinement in the $z$ direction is sufficiently
strong so that we can ignore the excitations in that direction,
and we will consider that the resulting potential in two dimensions is
isotropic and harmonic. The Hamiltonian describing this situation
in a frame rotating with the trap is: 
\begin{equation}
H=\frac{1}{2}
\sum_{i=1}^{N} \left(- \nabla_{i}^2 + r_{i}^2  -
2\frac{\Omega}{\omega}L_{iz}\right) + g \sum_{i<j}^{N}\delta
({\bf r_{i}} - {\bf r_{j}}), 
\label{H2D} 
\end{equation} 
with $L_{iz}$ being
the $z$ component of the angular momentum of the $i$--th atom, and
where we have used the trap energy, $\hbar \omega$, as the unit of
energy, and  $\ell=( \hbar/ m \omega  )^{1/2}$ as the unit of
length. The atoms are interacting via an effective contact
potential, 
with an interacting coupling constant $g$  related
to the $s$--wave scattering length, $a$, and to the localization
length in the $z$ direction, $\ell_{z}$, by $g=\sqrt{2/ \pi}a/ 
\ell_{z}$. In most parts of this paper we will also assume that
the interaction between the atoms is spin independent.
                            
Following our previous work \cite{Belen} we begin by writing the Hamiltonian 
(\ref{H2D}) in the following
form: 
\begin{equation}
H=H_{B}+H_{L}+ V.
\label{Hini}
\end{equation}
Here, $H_{B}=\sum_{i=1}^{N} -
\nabla_{i}^2/2 +r_{i}^2/2 - L_{iz}$ is the quantum Hall single
particle Hamiltonian, whose single particle energy levels are the
Landau levels equally spaced by the cyclotron energy
$2\hbar\omega$. The Hamiltonian $H_{L}=(1-\Omega/\omega)L_{z}$  is
proportional to the $z$ component of the total angular momentum,
$L_{z}=\sum_{i=1}^NL_{iz}$, and $V$ is the interaction term. 

From now on we consider the 
{\em rapid rotation limit}, in which the energy scales
characterizing Hamiltonians $H_{B}$ and $V$ are much larger than
the one corresponding to $H_{L}$. 
In this limit  the ground state and
elementary excitations of the system must lie on the subspace 
$\mathcal{F}$ of
common  zero energy eigenstates of  $H_{B}$ and $V$. 

We will show that, projected to the subspace $\mathcal{F}$,
the bosonic problem described
above  exactly maps
onto a problem of non-interacting fermions with $n$ internal levels.
We will first show that  any wave function in $\mathcal{F}$
is the product of a $\nu=1$ Laughlin state and a 
completely antisymmetric wave function in the lowest
Landau level. We will then see that this antisymmetric
wave function describes an effective system of non-interacting
fermions.

Let $\Psi$ be a wave function in $\mathcal{F}$.
In order to be  an eigenstate of 
$H_{B}$ with eigenvalue zero,  $\Psi$
must lie within the subspace spanned by tensor
products of single particle states in the
lowest Landau level  \cite{QHE}.
In the symmetric gauge these single particle states 
are labeled by the third component of the angular 
momentum $m$ and  the spin $\sigma$, and have the
following form: 
\begin{equation}
\varphi_{m\sigma}(\xi)=\frac{1}{\sqrt { \pi 2^{m+1}}} z^{m} 
e^{-\vert z \vert^{2}/4} \; \; \; u^{\sigma},
\label{sp}
\end{equation}
where $\xi=(z,u)$, $z=x+iy$, and $u^{\sigma}$ is a $s$--spinor
with components $u^{\sigma}(\beta)=\delta_{\sigma \beta}$,
for $\sigma, \beta=-s,\ldots ,s$.
It follows that $\Psi$   
must have the form:
\begin{eqnarray}
\Psi[\xi]= 
\mathop{
\sum_{m_{1},\ldots ,m_{N}}}_{\sigma_{1}, \ldots ,\sigma_{N}}
\hspace{-0.35cm}
\alpha_{m_{1},\ldots ,m_{N}}
\;
z_{1}^{m_{1}} \ldots z_{N}^{m_{N}}
\;
u_{1}^{\sigma_{1}} \ldots u_{N}^{\sigma_{N}},
\label{suma}
\end{eqnarray}
with $m_{i}=0, \ldots ,\infty$, $\sigma_{i}=-s, \ldots ,s$,
and
where for simplicity we have omitted the global
exponential factor $\prod_{k}e^{-\vert z_{k} \vert^{2}/4}$\footnote{
For the sake of short notation, in most part of this
paper we omit the exponential factor $\prod_{k}e^{-\vert z_{k} \vert^{2}/4}$,
as well as the normalization factors.}.

The wave function
$\Psi$ must be an eigenstate of $V$ with eigenvalue zero as well.
We will show that  the function $\prod _{i<j} (z_{i}-z_{j})$
is a factor of $\Psi$.
Let's choose any pair
of particles $i$ and $j$. The dependence of $\Psi$ on $z_{i}$
and $z_{j}$ can be reexpressed in terms of the relative and center
mass coordinates, $z_{ij}$, $Z_{ij}$, so that we can expand     
$\Psi[\xi]=\sum_{m} z_{ij}^{m}F_{m}$, where $F_{m}$ depends on $Z_{ij}$,
$u_{i}$, $u_{j}$,
and on the positions and spinors of all the other particles. 
In order
for $\Psi$ to be annihilated by the hard-core interaction $V$,
$F_{0}$ must be identically zero. It follows that $\forall i,j$,
$z_{ij}$ is
a factor of $\Psi$, so that            
\begin{equation}
\Psi[\xi] = \Phi[\xi]\prod _{i<j} (z_{i}-z_{j}),
\label{fact}
\end{equation} 
where $\Phi[\xi]$ is an analytic
function of $z_{i}$ for all $i$,
 and therefore it lies in the lowest Landau level.

The factor $\Pi[z]=
\prod_{i<j}  (z_{i}-z_{j})$ is proportional,
up to a normalization constant, to the $\nu =1$ Laughlin wave function.
Since this factor  is fully
antisymmetric, the function $\Phi$
 must be completely antisymmetric too, for $\Psi$ to be
a bosonic wave function.
The factorization (\ref{fact}) thus states that 
any
{\em bosonic} wave function
lying within the subspace $\mathcal{F}$
is completely specified by a {\em fermionic}
wave function in the lowest Landau level.
                   
We will derive now an effective Hamiltonian for the
fermionic wave function $\Phi$, in the sense that the spectrum
of this effective Hamiltonian will be identical
to the one of  Hamiltonian (\ref{Hini}), projected onto the
subspace $\mathcal{F}$.

Let $
\Psi_{n}=\Phi_{n}\Pi$
be an eigenstate
   of Hamiltonian (\ref{Hini})
with eigenvalue $E_{n}$.
This state satisfies
\begin{equation}
H\Psi_{n}=H_{L}\Psi_{n}=E_{n}\Psi_{n}.
\end{equation}
Since   $\Pi$ is an eigenfunction of $H_{L}$ with eigenvalue
$ (1-\Omega/\omega)N(N-1)/2$, we can write the above expression in the form:
\begin{eqnarray}
E_{n}\Phi_{n}\Pi &=& \left( H_{L}\Phi_{n}\right) \Pi+
\Phi_{n} \left( H_{L} \Pi \right) \\
&=&(1-\Omega/\omega)[ \left(  L_{z}+N(N-1)/2 \right) \Phi_{n} ]
\Pi.
\end{eqnarray}
We have then proved that if $\Psi_{n}$
is an eigenstate of Hamiltonian (\ref{Hini})
the corresponding fermionic state $\Phi_{n}$
is an eigenstate of the effective Hamiltonian
\begin{equation}
H_{eff}=(1-\Omega/\omega)\left[ L_{z}+N(N-1)/2  \right],
\label{heff}
\end{equation}
with the same eigenvalue.
This means that our problem of $N$ interacting bosons 
projected to the subspace $\mathcal{F}$
reduces to an effective problem 
of $N${ \em free} fermions in the lowest landau level, 
governed by the Hamiltonian (\ref{heff}).
The spectrum of the real bosonic system is identical
to the one of the effective free fermions, with the
same eigenvalues and the same density of states. The fermionic 
eigenstates are connected to the
bosonic ones by the transformation (\ref{fact}).
Note that this transformation is not unitary,
and therefore
an orthonormal set of fermionic states will be 
transform into a set of bosonic states that
will not be, in general, orthonormal.

The problem of free fermions in the lowest Landau level
with Hamiltonian  (\ref{heff}) can be easily solved.
The ground state is  a   Fermi sea, in which the effective
fermions fill up the single particle states
$\varphi_{m\sigma}$  in order of increasing angular
momentum $m$. Note that for each angular momentum $m$ we have
$n$ single particle degenerate states.
For a given number of particles $N$ we can write $N=kn+r$,
with $0 \le r <n$. This means  that we will have the first
$k$ single particle levels completely occupied and the $k+1$ level
partially occupied with $r$ fermions.
The ground state will have degeneracy
{\small $\left( \begin{array}{c} n\\r \end{array} \right) $}, corresponding
to the different choices for the internal states of the
last $r$  fermions.

If $\Phi_{FS}$ is the wave function corresponding
to the Fermi sea,
the bosonic ground state is then just:
\begin{equation}
\Psi[\xi]=\Phi_{FS}[\xi]\prod _{i<j} (z_{i}-z_{j}).
\end{equation}

Within this  effective fermionic picture it  is also very simple
to find the low-lying excitations of the system.
They are fermion-like excitations in  which an effective
fermion crosses the Fermi level, leaving a fermionic--hole
in the Fermi sea.

\section{Multicomponent atomic fractional quantum Hall liquids
and their anyons}
In this section we will study  the  case in
which the ratio $N/n$ is an integer number $k$, so that 
we have a non-degenerate ground state with $k$ completely occupied
fermionic levels. We will show that the effective fermions
form integer quantum Hall states at filling factor
$\nu=n$. The corresponding bosonic states are $\nu=n/(1+n)$ fractional
quantum Hall liquids with $1/n$--anyons.

\subsection{The $\nu=1/2$ Laughlin liquid}
We start out with  the trivial case in which $n=1$
so that there is no spin degeneracy.
The effective fermions will in this case occupy
the single particle states with angular momentum
$m=0, \ldots ,N-1$.  The Slater determinant
corresponding to this situation has the simple form:
\begin{equation}
\Phi_{1}[z]= \prod_{i<j}^{N}(z_{i}-z_{j}).
\end{equation}
This state is  the $\nu=1$ Laughlin state, the same as for electrons
in the quantum Hall effect.
The density of this state is nearly uniform, with
one fermion on average per unit area.
The corresponding bosonic state, 
\begin{equation}
\Psi[z]= \prod_{i<j}^{N}(z_{i}-z_{j})^{2},
\end{equation}
is
the $\nu=1/2$ Laughlin liquid that we studied in our
previous work \cite{Belen}. 
Starting from this liquid one  can create $1/2$--atomic--quasihole 
excitations with $1/2$--fractional statistics. The fractional statistical
phase of these quasiholes may be revealed directly in  a 
Ramsey-type interferometer \cite{Belen}.

\subsection{The $\nu=2/3$  liquid}

Before considering the general case of any number of
internal states, it is illuminating to study the
case of two internal states (and an  even number of atoms).
We will use the spin language, and say that
we have atoms with spin $s=1/2$ that can be either
up ($\uparrow$) or down ($\downarrow$). 

The effective fermions will fill up the single particle
states with angular momentum $m=0, \ldots ,N/2-1$ and
both spin states $\sigma = \uparrow, \downarrow$.
We will have $N/2$ fermions with spin up and 
$N/2$ fermions with spin down, and each group
will form a $\nu =1$ Laughlin state. 
The wave function corresponding to this situation is
\begin{eqnarray}
\Phi[\xi]=&\mathcal{A}&\left\{ \phi_{1}^{\uparrow}(\xi_{1}, 
\ldots ,\xi_{N/2})\right. \times \nonumber \\ 
&&\left. \phi_{1}^{\downarrow}
(\xi_{N/2+1}, \ldots ,\xi_{N})
\right\} ,
\label{nu2}
\end{eqnarray}
where
\begin{equation}
\phi_{1}^{\sigma}[\xi]= \prod_{i<j}^{N/2}(z_{i}-z_{j})u_{i}^{\sigma}
u_{j}^{\sigma}, \hspace{0.5cm} \sigma=\uparrow, \downarrow, 
\end{equation}
is  a $\nu =1$ Laughlin state of $N/2$ fermions in the spin 
state $\sigma$, and the  operator $\mathcal{A}$ antisymmetrizes over all  
possible ways of distributing the fermions in two groups
of $N/2$ fermions.
The state (\ref{nu2}) is a $\nu=2$ quantum Hall state,
made out of two $\nu=1$ states, one with spin up and the
other with spin down. The density profile of this state
is nearly flat in the bulk, with two fermions per unit area on average,
one with spin up and the other with spin down.

The corresponding bosonic state is obtained from 
the wave function (\ref{nu2}) multiplying by the factor
$\prod^{N} (z_{i}-z_{j})$. To gain insight about the properties of this
bosonic state it is very convenient to use the 
language of the quantum Hall effect, that relates 
angular momenta, filling factors and densities \cite{QHE}.
In general, given an incompressible liquid in the lowest
Landau level,  its filling factor $\nu$ is related to
its total angular momentum by $L\sim N^{2}/2\nu$.
Furthermore, the  liquid has a uniform density in the bulk,
with $\nu$ particles per unit area on the average.

Multiplication by the factor $\Pi$ increases the total
angular momentum $L_{F}$ of the original fermionic state, so that
the resulting bosonic state has a total angular momentum
\begin{equation}
L=L_{F} + \frac{N(N-1)}{2}.
\end{equation}
This means that the new bosonic state will have a filling factor
\begin{equation}
\nu=\frac{\nu_{F}}{1+\nu_{F}}.
\label{nunuf}
\end{equation}
As $\nu_{F}=2$ we have that the bosonic state is
a $\nu=2/3$ fractional Hall liquid.
This fractional liquid  is made out of two $\nu=1/3$ liquids, one with
spin up and the other with spin down.
The average density in this state is of $2/3$ atoms
per unit area, $1/3$ with spin up, and $1/3$ with spin 
down. Roughly speaking, we can say that  one bosonic atom is  made out of 
three effective fermions, or that one fermion
corresponds to $1/3$ atom. This counting  will be very
useful to study the anyonic excitations in the fermionic
picture.

\subsection{The $\nu=n/(1+n)$ multispinor liquids}
After the above discussion the generalization to
any number of internal states is straightforward.
The effective fermions will fill up the single
particle states with angular momenta $m=0, \ldots, N/n-1$,
and spin $\sigma=-s, \ldots ,s$.
We will have $n$ groups of  $N/n$ fermions, one for each
spin state, and each
group will form a $\nu=1$ polarized liquid of the form
\begin{equation}
\phi_{1}^{\sigma}[\xi]=\prod_{i<j}^{N/n}
(z_{i}-z_{j})u_{i}^{\sigma}u_{j}^{\sigma}.
\end{equation}
The fermionic ground state is the product of the
wave functions of the different groups, antisymmetrized
over all possible ways of distributing $N$ particles
in  groups of 
$N/n$ particles:
\begin{equation}
\Phi[\xi]=\mathcal{A} \left\{  \prod_{\sigma}\phi_{1}^{\sigma}[\xi] \right\}
\label{nun}
\end{equation}

The state (\ref{nun}) is a $\nu=n$ quantum Hall state,
consisting of $n$ polarized $\nu=1$ liquids,  one
for each possible spin state. In this state we have
$n$ fermions per unit area, one in each of the spin states.

The relation $(\ref{nunuf})$ tells us that the corresponding 
bosonic state will be a fractional state with filling
 factor $\nu=n/(1+n)$. This liquid is made out of 
$n$ polarized $\nu_{\sigma}=1/(1+n)$ fractional liquids,
one for each spin state $\sigma$.
We will have $n/(1+n)$ atoms per unit area, $1/(1+n)$ per
each spin state.  

By increasing  the spin degeneracy $n=1, 2, 3, \ldots$,
one can drive the atomic system into fractional
quantum Hall states in the sequence
 $\nu=1/2, 2/3, 3/4, \ldots   $.

\subsection{Anyons}

We will show now how to create anyons starting from the fractional
atomic quantum liquids presented in the preceeding subsection.
Following the ideas of our previous work \cite{Belen} we insert a laser
localized (within  an area $\ell^{2}$) at some position  $\eta$.
We require $|\eta|$ to be much smaller than the size of the quantum Hall state
$(\sim \nu^{-1}\sqrt {N-1})$, so that the border effects are negligible.
We will also assume that the laser only couples to atoms with spin
$\sigma_{0}$. The presence of the laser can be described by
a localized repulsive potential, so that the new Hamiltonian
for the bosonic atoms  can be approximated by:
\begin{equation}
H^{\eta \sigma_{0}}=H+V_{0}\sum_{i}\delta(z_{i}-\eta)P_{i}
^{\sigma_{0}},
\label{hdelb}
\end{equation} 
where the operator $P_{i}^{\sigma_{0}}$ projects the spinor 
of the $i$--th particle on the state with  spinor
$u_{i}^{\sigma_{0}}$. In the same way as we did in the previous
section we can project Hamiltonian (\ref{hdelb}) onto the subspace
$\mathcal{F}$ of wave functions of the form $\Psi=\Phi
\prod_{i<j}(z_{i}-z_{j})$, and
derive an effective Hamiltonian for the fermionic wave function
$\Phi$, which has now the form \footnote{Strictly speaking one obtains the
Hamiltonian (\ref{hdel}) by defining a new scalar
product in the fermionic Hilbert space of the form
$\langle \Phi_{1}\vert \Phi_{2} \rangle=\int d\mu^{2} \Phi_{1}^{*}
\Phi_{2}$,
 where $d\mu^{2}=\prod_{i<j}\vert z_{i}-z_{j}\vert ^{2}dxdy$.}: 
\begin{equation}
H^{\eta \sigma_{0}}_{eff}=H_{eff}+V_{0}\sum_{i}\delta(z_{i}-\eta)P_{i}
^{\sigma_{0}}.
\label{hdel}
\end{equation}  
We will find the ground state of the fermionic Hamiltonian
(\ref{hdel}) and show that the corresponding bosonic state
is a fractional quasihole state localized at position $\eta$.
We consider the limit in which the intensity of the laser 
is much larger that the energy scale characterizing Hamiltonian
$H_{eff}$.
In this limit the new ground state $\Phi_{\eta \sigma_{0}}$
        must  be a zero eigenstate
of the potential created by the laser. It follows that, 
whenever the $i$--th particle is in the spin state
$\sigma_{0}$, 
$(z_{i}
-\eta)$ must be a factor of $\Phi_{\eta \sigma_{0}}$. 

All groups with $\sigma \neq \sigma_{0}$ will remain in  a  
$\nu=1$ Laughlin state with wave function $\phi^{\sigma}[z]$.
But the  group with $\sigma = \sigma_{0}$ will form  the 
quasihole state
\begin{equation}
\phi_{\eta}^{\sigma_{0}}[\xi]=\prod_{i=1}^{N/s}(z_{i}-\eta)
\phi^{\sigma_{0}}[\xi].
\end{equation}

The resulting state is a $\nu_{F}=n$ fermionic
liquid in which  one fermion with spin $\sigma_{0}$ has been 
removed at position $\eta$.
As each boson is made out of $n+1$ fermions,
the corresponding bosonic state will be  a $\nu=n/(1+n)$
liquid in which a fraction $1/(1+n)$ of an atom is missing
in the component with spin $\sigma_{0}$.
As it follows from the theory of fractional quantum Hall effect,
these atomic fractional quasiholes
are anyons with $1/(1+n)$ fractional statistics.

In the analysis above we have considered a situation in 
which the laser only couples to atoms in a particular internal state
$\sigma_{0}$. But we can also imagine situations in which the
laser coupling affects  different internal states.  
By selecting the number of internal states coupled to the laser
we can create anyons with any fractional statistics of the form
$p/(1+n)$, with $1 \leq p \leq n$. 

The fractional statistics of the anyons we have described
may be  directly proved in an  experiment similar to the one we proposed
for detecting  $\frac{1}{2}$--anyons \cite{Belen}. 
Suppose that we want to detect $\frac {1}{3}$--anyons.
Then we will first prepare a system of bosons 
with $2$ internal states, up 
and down,
in a $\nu=2/3$ state. We focus then a laser, that only couples to 
atoms with spin up  (or down), at position $\eta_{1}$,
and increase its intensity until a single $\frac{1}{3}$--quasihole is
created. Keeping constant the intensity of this laser we then adiabatically insert another laser at position $\eta_{2}$, far enough from $\eta_{1}$,
affecting also only atoms with
spin up (or down).
The system then evolves from the one--quasihole state,
$\Psi_{\eta_{1}}$, to a two--quasihole state, $\Psi_{\eta_{1},\eta_{2}}$.
The crucial point is that at a certain point of the evolution
the system reaches a superposition state $\Psi \sim \Psi_{\eta_{1}}+
\Psi_{\eta_{1},\eta_{2}}$.
This is the superposition we need to test the statistical angle.
If we adiabatically move the laser at position $\eta_{1}$
along a path enclosing position $\eta_{2}$, the evolved state
at the end of the process will be 
\begin{equation}
\Psi^{\prime}\sim  \Psi_{\eta_{1}} + e^{i2\pi/3} \Psi_{\eta_{1},\eta_{2}},
\end{equation}
where the relative phase, $e^{i2\pi/3}$, reflects the statistical phase.
This phase may be detected in a Ramsey-type interferometer in which
the evolution of the initial superposition state, $\Psi$, is split in 
three different ways, in which the laser at $\eta_{1}$ performs
respectively, one, two, and three loops enclosing $\eta_{2}$.
If after performing the loops we continue the evolution  of the system,
one can check that the system will only evolve to the two-quasihole
state, $\Psi_{\eta_{1},\eta_{2}}$, in the case in which $3$ loops
were performed. This $3$ 
reflects the $\frac{1}{3}$--statistics of the
quasiparticles.

\section{The fermionic bosons}
In this section we give a  special attention to the case in which
the degeneracy of spin equals the number of atoms in the system.
Such a situation is a realistic possibility with a system of
small number of atoms.
In this case the effective fermions will occupy the single 
particle states with $m=0$ and  spin  $\sigma=1, \ldots  ,N$.
The corresponding Slater determinant is a pure spinor function,
with no spatial dependence
\begin{equation}
\Phi[\xi]=\chi [u]=\sum_{i_{1}, \ldots ,i_{N}} \epsilon_{i_{1}, \ldots ,i_{N}}
u^{1}_{i_{1}} \ldots u^{N}_{i_{N}},
\end{equation} 
so that the bosonic ground state
\begin{equation}
\Psi[\xi]=\chi[u]  \prod_{i<j}(z_{i}-z_{j})
\label{fer}
\end{equation}
is the product of  a completely antisymmetric spin state
and a $\nu=1$ Laughlin state.
We note that in the above wave function
the spin and orbital degrees of freedom are
completely decoupled .
The orbital wave function is antisymmetric, 
so that as long as the spin degrees of freedom
are not excited the behaviour of the system
will be fermionic. Any property concerning 
orbital degrees of freedom that were
experimentally measured would behave as if we would
have fermions in the system, even though our system
is composed of bosons.

Suppose, for example, that we
insert a laser localized at a certain position
$\eta$, and  that the laser couples to all atoms,
no matter the internal state. As the intensity of the
laser is increased the system will evolve from state
(\ref{fer}) to the quasihole state:
\begin{equation}
\Psi_{\eta}[\xi]=\chi[u]\prod_{i}(z_{i}-\eta)  \prod_{i<j}(z_{i}-z_{j}).
\end{equation}
This quasihole state is a $\nu=1$ state in which one particle
has been removed at position $\eta$, therefore, the relation
between filling factors and fractional statistics states that
we have created an  anyon with statistics $1$; that is, a fermion.

\section{Numerical simulations}
In this section we illustrate the results presented in
the previous sections  with
some numerical exact diagonalizations for   $N=4,6,$ and $8$ bosons.
In our calculations we restrict ourselves to single-particle
states in the lowest Landau level. This approximation is valid 
in the limit in which the energy separation between Landau levels, $2\omega$,
is much larger than all the energies avaliable.

We note that Hamiltonian (\ref{Hini}) is invariant 
under rotations around the $z$ axis,
and also under global spin rotations. These symmetries allow us to
 diagonalize  it  within
 subspaces of fixed $z$ component
of the total angular momentum, $L$,  fixed total spin, $S$, and fixed
$z$ component of the spin, $S_{z}$.

\begin{figure}[!]
\begin{center}                                                
\includegraphics[height=7cm,width=7cm]{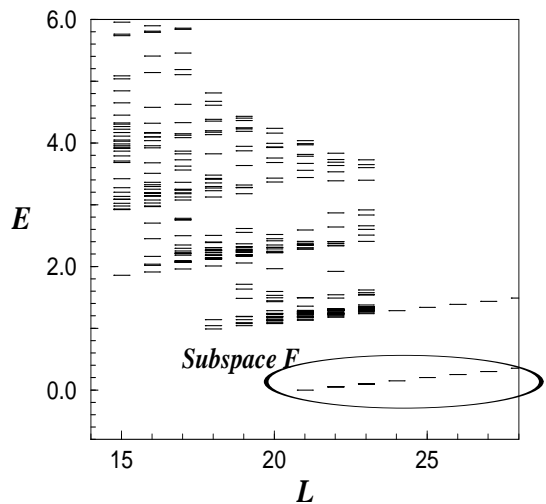}       
\end{center}
\caption{Eigenvalues of Hamiltonian(\ref{Hini})  for $N=6$ bosons
and $n=2$ internal levels, as a function of the total
angular momentum $L$. For illustration purposes we chose
$(1-\Omega/\omega)/g=0.01$. The energy is measured in 
units of $g \hbar     \omega$, and all energies are 
shifted by $\Delta E=-2.1g \hbar     \omega$.
For total angular momenta larger than $L=23$ the dimension
of the subspace is too large and we can not diagonalize the
Hamiltonian exactly. We have used a variational approach to 
obtain the ground state and the first excited states.} 
\label{fig1}
\end{figure}

Figure \ref{fig1}   shows the eigenvalues of Hamiltonian
$(\ref{Hini})$ for a system of
$N=6$ bosons with two internal levels, in the rapid rotation
limit. We see that
there is a branch of states well separated
by a gap from the rest of the spectrum. As
long as the temperature is smaller than this gap,
the system will remain within this subspace. All the states
in this subspace lie within the lowest Landau level and
have zero interaction energy. They form what we have called
              subspace $\mathcal{F}$.

\begin{figure}[!]
\begin{center}                                                
\includegraphics[height=5.5cm,width=7cm]{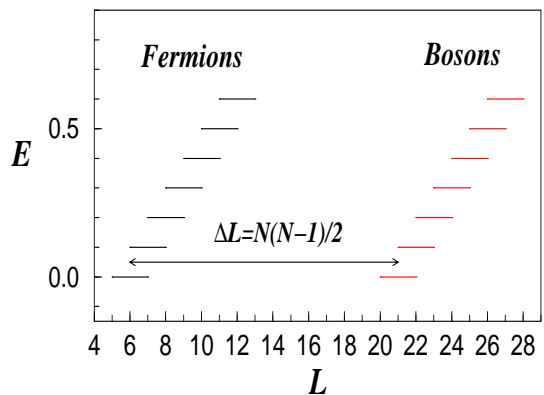}       
\end{center}
\caption{Eigenvalues of Hamiltonian (\ref{Hini}) for $N=6$ bosons
projected to the subspace $\mathcal{F}$ (right), and
eigenvalues of Hamiltonian (\ref{heff})  for $N=6$ fermions (left),
as a function of the total
angular momentum $L$.
Both spectra are identical except for a shift in the angular
momentum $\Delta L=N(N-1)/2$.
The data shown correspond to 
$(1-\Omega/\omega)/g=0.01$. The energy is measured in 
units of $g \hbar     \omega$, and all energies are 
shifted by $\Delta E=-2.1g \hbar     \omega$.}
\label{fig2}
\end{figure}

\begin{figure}[!]
\begin{center}                                                
\includegraphics[height=7cm,width=7cm]{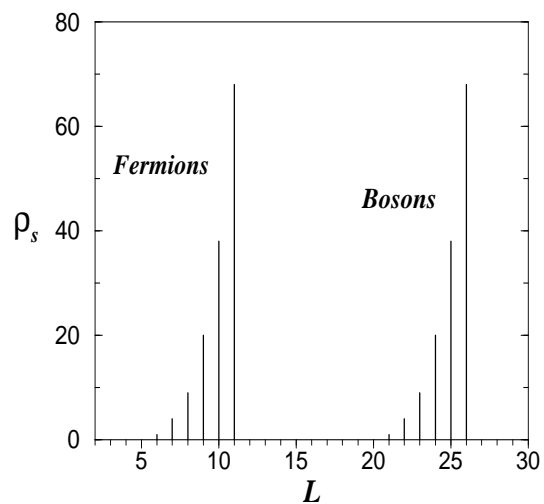}       
\end{center}
\caption{Density of states for $N=6$ bosons with $n=2$ internal levels
in the subspace $\mathcal{F}$, as a function
of the total angular momentum $L$. We have also plotted
the density of states for a system of $N=6$ fermions
and $n=2$ internal levels with Hamiltonian (\ref{heff}).}\label{fig3} 
\end{figure}

In order to check the fermionization scheme we have diagonalized
the effective Hamiltonian (\ref{heff})
 for a system of $N=6$ fermions with spin  
$s=1/2$. Figure \ref{fig2} shows that the bosonic spectrum
projected to the subspace  $\mathcal{F}$ is identical to that
of the free fermions, except for a shift in the angular momentum
of the eigenstates, $\Delta L=N(N-1)/2$. This is precisely the
angular momentum of the $\nu=1$ Laughlin state, that connects
fermionic and bosonic states.
For a given angular momentum, there are many degenerate
states within the subspace $\mathcal{F}$.  
This density 
of states is exactly reproduced by the system of non-interacting
fermions (Figure \ref{fig3}),
exept again,
for the shift in angular momentum.

\begin{figure}[!]
\begin{center}                                                
\includegraphics[height=7cm,width=7cm]{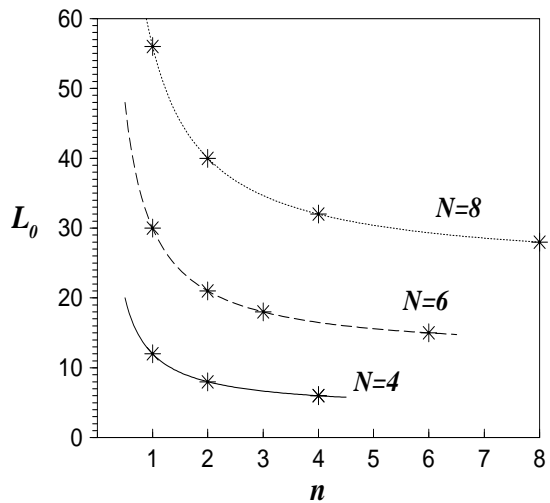}       
\end{center}
\caption{Total angular momentum $L_{0}$ of the ground state
for a system with $N=4,6,8$ bosons, as a function of the number of
internal levels $n$. We have plotted the numerical values (stars)
together with the curve given by equation (\ref{curve}). }\label{fig4}
\end{figure}

We have studied the ground state properties of a system
of $N=4,6,8$ bosons, and $n$ internal states, for the cases
in which $N/n$ is an integer number.
We find (Figure \ref{fig4}) that the total angular momentum $L_{0}$
of the ground state is always equal to the one of the
Fermi sea formed by the fermions, plus the angular 
momentum of the $\nu=1$ Laughlin state:
\begin{equation}
L_{0}=N(N/n-1)+N(N-1)/2.
\label{curve}
\end{equation}

\begin{figure}[!]
\begin{center}                                                
\includegraphics[height=7cm,width=7cm]{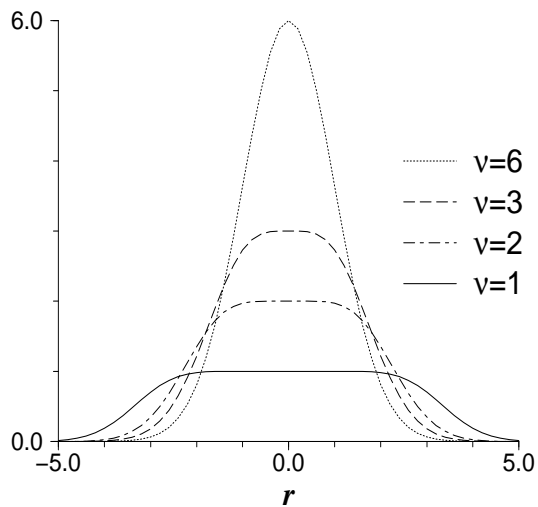}       
\end{center}
\caption{Density profile of the ground state of a system of
$N=6$ fermions with Hamiltonian (\ref{heff}). The different
curves correspond to different values of the number of
internal states $n$. The unit of density is $1/(2\pi\ell ^{2})$,
and the unit of length is $\ell$.}\label{fig5}
\end{figure}

\begin{figure}[!]
\begin{center}                                                
\includegraphics[height=6cm,width=7cm]{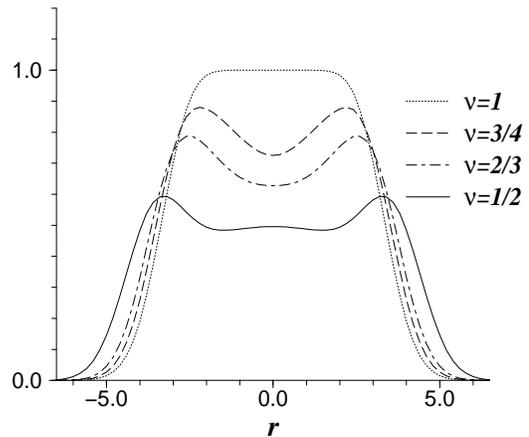}       
\end{center}
\caption{Density profile of the ground state of a system of
$N=6$ bosons with Hamiltonian (\ref{Hini}), in the limit
of rapid rotation. The different
curves correspond to different values of the number of
internal states $n$. The unit of density is $1/(2\pi\ell ^{2})$,
and the unit of length is $\ell$.}
\label{fig6} 
\end{figure}

We have checked that the effective fermions form integer
quantum Hall states at filling factor $\nu=n$. The density 
profile of these states is nearly flat in the bulk, with
$n$ fermions per unit of area (Figure \ref{fig5}).
The corresponding bosonic states are fractional quantum
Hall liquids at filling factor $\nu=n/(n+1)$. The density
profile of these states is also nearly flat in the bulk, with
$n/(1+n)$ bosons per unit area (Figure \ref{fig6}).

\begin{figure}[!]
\begin{center}                                                
\includegraphics[height=6cm,width=7cm]{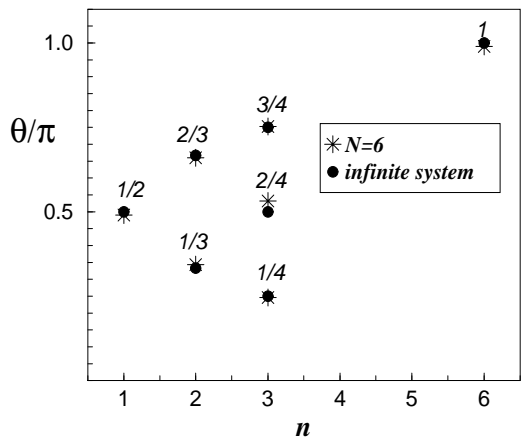}       
\end{center}
\caption{Statistical angle of the quasiholes created in
a system of $N=6$ bosons, as a function of the number of
internal states. We have plotted the numerical results (stars)
together with the ones predicted for an infinite system (dots).
The different points for each number of internal states
correspond to different values of the number of internal states
the laser is coupled to. }\label{fig7}
\end{figure}

We have also checked that, starting from these fractional
liquids,  anyons with $p/(1+n)$ statistics may be created.
We consider first the following Hamiltonian,
\begin{equation}
H_{0}=H+V_{1}\sum_{i}\delta(z_{i}-\eta_{1})P_{i}
^{\sigma_{1} \ldots \sigma_{p}},
\label{hdeld}
\end{equation} 
which includes the presence of a laser localized at  position
$\eta_{1}$,  affecting $p$ different spin states.
We have diagonalized Hamiltonian (\ref{hdeld}) within
the subspace $\mathcal{F}$ for $N=6$
bosons, $n=1,2,3,6$, and $1 \leq p \leq n$. When the laser
power is sufficiently large we find that the ground state
is a one-quasihole state, $\Psi_{\eta_{1}}$, in which a certain amount of
atoms is missing in the components with spin states 
$\sigma_{1}, \ldots ,\sigma_{p}$.
We have also  diagonalized a Hamiltonian describing the
system in the presence of two lasers localized at positions
$\eta_{1}$
and $\eta_{2}$, coupled to the same $p$ internal states.
The resulting ground state, $\Psi_{\eta_{1},\eta_{2}}$,  
consists of two identical quasiholes
located at positions $\eta_{1}$
and $\eta_{2}$. 
Having the states $\Psi_{\eta_{1}}$ and $\Psi_{\eta_{1},\eta_{2}}$ we can 
check very easily what is the statistics of the quasiholes
we have created.
Suppose that we adiabatically drive the laser at position
$\eta_{1}$ along a path enclosing position $ \eta_{2}$.
At the end of the process the states $\Psi_{\eta_{1}}$ and $\Psi_{\eta_{1},\eta_{2}}$
will pick up Berry phases given, respectively, by,
\begin{eqnarray}
\gamma_{1}&=&2\pi\int_{A}dx dy |\Psi_{\eta_{1}} |^{2} \nonumber\\
\gamma_{2}&=&2\pi\int_{A}dx dy |\Psi_{\eta_{1},\eta_{2} } |^{2},
\label{berry}
\end{eqnarray}
where the integrals are performed over the area $A$ enclosed
by the path described by the laser. The equations (\ref{berry}) for the 
Berry phases can be easily derived, taking into account
that the states $\Psi_{\eta_{1}}$ and $\Psi_{\eta_{1},\eta_{2}}$ are made out of fractional
Laughlin quasihole states \cite{anyons}.
The difference between the Berry phases $\gamma_{2}-\gamma_{1}$ will reflect the extra phase
that the quasihole at $\eta_{1}$ picks up because of the presence of the other
quasihole at $\eta_{2}$. Since the closed loop we have performed
is equivalent to two consecutive interchanges of the quasiholes we have that the statistics of the
quasiparticles is given by the angle

\begin{equation}
\theta=\frac {\gamma_{2}-\gamma_{1}}{2}.
\label{teta}
\end{equation}
We have calculated expression (\ref{teta}) for quasihole states
corresponding to $N=6$ bosons and different 
numbers of internal states $n=1,2,3,6$. The statistical phases we obtain
(Figure \ref{fig7})
are in excellent agreement with the fractions $p/(1+n)$, even though
the finite size of the system.

\section{Experimental conditions}

In this section we discuss the set of conditions that 
 a system of bosonic atoms must fulfill  in order to  observe
 the fractional 
quantum Hall liquids and  the anyons we have described
in the preceeding sections.

First of all we need a system with degenerate internal levels, such that
the interaction between the atoms does not depend on the internal state.
We  have also made a two-dimensional approximation,
so that the atoms must be confined in a trap with  $\ell \ll \ell_{z}$.
For the ground state  to be a fractional quantum Hall
liquid the system must be in the rapid rotation regime. From
the numerical diagonalizations we can calculate the critical 
frequency of rotation at which the system will be driven to a correlated
quantum Hall liquid. For a system of $N=6$ particles we find that
 the ground state of the system becomes a $\nu=1/2,2/3,3/4$ liquid
for a frequency of rotation $\Omega/\omega\geq 0.80, 0.62, 0.38$,
 respectively.
These data show 
how  the critical frequency of rotation
decreases as the number of internal states is increased, so that
 the observation of the fractional states
 $\nu=n/(1+n)$ requires
less faster rotating traps, as we approximate $\nu=1$.
For an arbitrary number of particles $N$ we can estimate what is
the condition that the frequency of rotation must fulfill in order
for the a state $\nu=n/(1+n)$ to be the ground state.
The state $\nu=n/(1+n)$ has a total angular momentum 
$L \sim \nu^{-1} N(N-1)$ (decreasing as $\nu \rightarrow 1$),
 and thus an angular momentum energy per
particle $e_{L} \lesssim \nu^{-1}(1-\Omega/\omega)(N-1)$. 
In order to be in the rapid rotation regime  we  need $e_{L}$ 
to be much  
smaller than both the trap energy and the typical interaction     
energy, so that the conditions 
$(1-\Omega/\omega)(N-1)\ll \nu, \        
\nu g /4\pi$
must be fulfill. 

For creating the anyons the conditions required are much more
demanding. We have first to focus the lasers within a distance
$\sim \ell$. 
For a localization length $\ell \sim 1\mu m$
this implies an upper limit for the trap frequency of $\sim 1kHz$.
As well, the frequency of rotation needed to create the anyonic
excitations is larger that the critical frequency to observe the
fractional quantum Hall liquids, since the total angular momentum
of the quasihole states is $2N$ units of angular momentum larger.
For a system with $N=6$ particles  anyons
with $1/2,1/3,1/4$--statistics may be created for $\Omega/\omega \geq 
0.91, 0.81, 0.68$, respectively.

Finally , the most restrictive condition to observe the
fractional quantum Hall states and their anyons is the temperature. 
In order to freeze out the      
excitations we need $kT/\hbar\omega \ll (1-\Omega/\omega)$.
For a  system of $N=6$ particles the observation of the states
$\nu=1/2,3/4,3/4$ requires temperatures $kT/\hbar\omega \ll 0.20,
0.38, 0.62$.

\begin{section}{The paired superconducting states}
In this section we  study a situation in which the ground
state of the atomic system is a paired state of effective fermions.
We consider a system with and even number of atoms and $n=2$ internal
levels, up ($\uparrow$), and down ($\downarrow$), and we assume
that the interaction  between atoms in the same spin state is
much larger than the interaction between atoms with different
spin states. In this case the interaction term in Hamiltonian
(\ref{Hini}) can be approximated by
\begin{equation}   
V^{\parallel}=g\sum_{i<j}^{N}
 \delta(z_{i}-z_{j})\left( P_{ij}^{\uparrow \uparrow}+
P_{ij}^{\downarrow \downarrow} \right),
\label{upup}
\end{equation}    
where the operator $P_{ij}^{\sigma \sigma}$ projects the spin state
of the pair of particles $ij$ on the state $\sigma \sigma$.

We will show that, in the limit of rapid rotation, the ground
state of the system is the product of a $\nu=1$ Laughlin state
and a completely antisymmetric wave function.
This antisymmetric wave function has the structure of a BCS state
of fermions with p-wave pairing.

From the results of  section II it follows that in the rapid rotation
limit the ground state, $\Psi$,  of the system must be both an
analytic function of the $z_{i}$'s and an eigenstate
of the interaction (\ref{upup}) with eigenvalue $0$.
Let's choose any pair of particles $i$ and $j$.
For $\Psi$ to be annihilated by the interaction (\ref{upup})
we have two possibilities. Either the particles $i$ and $j$ are
in a different spin state, or the relative angular momentum of
the pair is larger than $0$. It follows that
either the spin triplet  $(\uparrow_{i}\downarrow_{j}+
\downarrow_{i}\uparrow_{j})$ or the factor 
$(z_{i}-z_{j})$  is a factor of $\Psi$.
How many factors of the form $(z_{i}-z_{j})$ and how many 
spin triplets we will have in the ground  state will be the choice
that costs the minimal amount of angular momentum energy, thus,
we will have as many triplet states as possible.
Note that the maximum number of pairs that we can arrange 
in triplets is precisely $N/2$, corresponding to a certain pairing
of the particles. 
It follows that the ground state has the form
\begin{equation}
\Psi=Pf  \left(\frac {\uparrow_{i}\downarrow_{j}+
\downarrow_{i}\uparrow_{j}}{z_{i}-z_{j}}\right)   \prod _{i<j} (z_{i}-z_{j}),
\label{pfaff}
\end{equation}
where the Pfaffian $Pf$ is defined by
\begin{equation}
Pf (M_{ij})= \mathcal{A}(M_{12}M_{34}\ldots M_{N-1,N}),
\end{equation}
where $M_{ij}$ are the elements of an antisymmetric matrix and $\mathcal{A}$
denotes the operation of antisymmetrization, normalized such that each 
distinct term appears one with coefficient 1.

The  state (\ref{pfaff}) is the product of a $\nu=1$ Laughlin
state, in which  the factor $(z_{i}-z_{j})$ appears for all pairs
of particles $ij$, and a Pfaffian factor. The Pfaffian factor acts
in the following way. It chooses a pairing of the particles, and
for all pairs $ij$ within the pairing it removes the factor
 $(z_{i}-z_{j})$ 
and substitutes it by a triplet state ($\uparrow_{i}\downarrow_{j}+
\downarrow_{i}\uparrow_{j}$). In this way the system minimizes its total
angular momentum, remaining in the subspace of zero interaction eigenstates.

Since the Pfaffian factor is completely antisymmetric we can interpret
it as the wave function of a system of $N$ effective fermions, in the
same way that we did in section II. In this case, however, the fermionic
wave function is not an analytic function of the $z_{i}$'s, and, therefore
it does not lie within the lowest Landau level. Such a Pfaffian state 
has been shown to be the ground state solution of a BCS theory for fermions
with p-wave triplet pairing \cite{Read}. With this result  we can say that the state
(\ref{pfaff}) is a superconducting state of effective fermions.
It would very interesting to derive an effective Hamiltonian
for the fermionic wave function, in a similar way
as we did in section II. This effective 
Hamiltonian should  describe in this case fermions
with attractive interactions between particles with different 
spin states. Derivation of this Hamiltonian
will be discussed elsewhere.

\section{Conclusions}

In conclusion, we have shown how by rapidly rotating
a gas of ultra cold bosonic atoms, we can drive the system into
a strong interacting regime in which novel correlated
ground states and excitations appear.

We have developed a theory that exactly maps the
problem of interacting bosons in the rapid rotation
regime to a problem of non--interacting fermions.
This fermionization scheme  allows us to find the exact ground state
and  low-lying excitations of the system. 
For the case of a system with no internal levels
we recover the results of our previous work, with a $\frac{1}{2}$--Laughlin
liquid as the ground state, and $1/2$ anyonic excitations. 
In the presence of  $n$  internal levels
and when the ratio $N/n$ is an integer number, the ground
state of the effective fermions is a $\nu=n$ integer quantum Hall state.
The corresponding bosonic ground state is a fractional
multicomponent quantum Hall  liquid with filling factor $\nu=n/(1+n)$.
This fractional liquid is a highly correlated state made out of $n$ copies of a
$1/(1+n)$--Laughlin liquid, one for each internal state.
Starting from these liquids we have shown how to create
$p/(1+n)$--anyons, by focusing lasers at the desired positions,
coupled to $p$ different spin states.
The fractional statistics of these anyons may be directly tested
in a Ramsey-type interferometer similar to the one
we proposed to detect $1/2$ anyons.

The experimental observation of the fractional quantum Hall
liquids and the anyons described in this work requires small
temperatures, small number of atoms, and very fast rotating
traps. The conditions required are relaxed as the number of
internal levels  increases.
A possible scenario for the observation of these entangled
states is a two--dimensional rotating optical lattice.
As already shown in the recent experiment with a one-dimensional optical
lattice it is possible to reach a regime in which a small fixed number of atoms
is confined in each well of the lattice.
Furthermore, since one has many identical copies of
the  system,  the experimental signals are highly magnified.

\section*{Acknowledgments}
We thank I. Bloch and M. Greiner for discussions. B. P.
is supported  by grant HPMF-CT-2000-01045   (Marie Curie fellowship)
of the European Community.
\end{section}


\begin{thebibliography}{99}                                       
\bibitem{BEC}                                                     
M. H. Anderson,                                        
J. R. Ensher, M. R. Matthews, C. E. Wieman, E. A. Cornell,            
Science 269, 198 (1995); K. B. Davis, M. O. Mewes,
M. R. Andrews, N. J. van Druten, D. S. Durfee,
D. M. Kurn, W. Ketterle,  Phys.           
Rev. Lett. {\bf 75}, 3969 (1995); C. C. Bradley, C. A. Sackett, R.           
G. Hulet, Phys. Rev. Lett. {\bf 78}, 985 (1997).                   
\bibitem{Stringari}
F. Dalfovo, S. Giorgini, L. P. Pitaevskii, S. Stringari, Rev. Mod. Phys. 
{\bf 71}, 463-512 (1999).
\bibitem{Feshbach}J. Stenger, S. Inouye, M. R. Andrews,
 H.-J. Miesner, D. M. Stamper-Kurn,  W. Ketterle,
 Phys. Rev. Lett. {\bf 82}, 2422 (1999) 
\bibitem{Sorensen}A. S{\o}rensen, L.-M. Duan, J. I. Cirac,
P. Zoller, Nature {\bf 409}, 63 (2001).
\bibitem{Dieter} D. Jaksch, C. Bruder, J. I. Cirac, C. W. Gardiner, P. Zoller,
 Phys. Rev. Lett. {\bf 81}, 3108 (1998); E. Demler, F. Zhou,
cond-mat/0104409.
\bibitem{Bloch} M. Greiner, O. Mandel, T. Esslinger, T. W. H\"{a}nsch,
I. Bloch, Nature {\bf 415}, 39 (2002)
\bibitem{vortices}K. W. Madison {\em et al.}, Phys. Rev. Lett. {\bf
84}, 806 (2000); J. R. Abo-Shaeer, C.  Raman, J. M.  Vogels, W. Ketterle,
Science {\bf 292}, 476 (2001);
D. L. Feder, A. A. Svidzinsky, A. L. Fetter,  C. W. Clark,
Phys. Rev. Lett. 86, 564 (2001).
 D. A. Butts, D. S. Rokhsar, Nature {\bf 397} 327,
(1999).
\bibitem{otros}N. R. Cooper, N. K. Wilkin,  J. M. F. Gunn,
 Phys. Rev. Lett. 87, 120405 (2001); N. K. Wilkin, J. M. F. Gunn,
 Phys. Rev. Lett. 84, 6 (2000); J. Sinova, C. B. Hanna,
A. H. MacDonald, cond-mat/0201020;
A. D. Jackson, G. M. Kavoulakis, B. Mottelson,  S. M. Reimann,
 Phys. Rev. Lett. 86, 945 (2001).
 

\bibitem{Belen} B. Paredes, P. Fedichev, J. I. Cirac, P. Zoller,
Phys. Rev. Lett. {\bf 87}, 10402 (2001)
\bibitem{QHE}T. Chakraborty and P. Pietil\"ainen in
{\em The Quantum Hall Effects: Fractional and Integral}
(Springer-Verlag, Berlin, 1995); A. H. MacDonald, cond-mat/9410047.
\bibitem{Laughlin}
R. B. Laughlin, 
Phys. Rev. Lett. {\bf 50}, 1395 (1983).
\bibitem{anyons}
D. Arovas, J. R. Schrieffer, F. Wilczek, Phys. Rev. Lett. {\bf
53}, 722 (1984)
\bibitem{Read}
N. Read, cond-mat/0011338; D. Green, N. Read,
Phys. Rev. B {\bf 61}, 10267 (2000)                                                                                                                                                        
\end{thebibliography}
\end{document}